\documentclass[twocolumn,prl,aps,superscriptaddress,showpacs]{revtex4-1}

\usepackage{mathptmx}
\usepackage{subfigure}
\usepackage{dcolumn}
\usepackage{amsmath,amssymb}
\usepackage{bm}
\usepackage{color}
\usepackage{latexsym}
\usepackage{epstopdf}
\usepackage{color}
\usepackage[english]{babel}
\usepackage{latexsym}

\usepackage{psfrag,graphicx} 
\usepackage{epsf} 
\usepackage{subfigure} 
\usepackage{amsmath} 
\usepackage{amssymb} 
\usepackage{amsfonts}
\usepackage{bm}
\usepackage{natbib}
\usepackage{epstopdf}
\usepackage{appendix}

\definecolor{mygrey}{gray}{0.35}
\definecolor{myblue}{rgb}{0.2,0.2,0.8}
\definecolor{myzard}{cmyk}{0,0,0.05,0}
\definecolor{mywhite}{rgb}{1,1,1}
\definecolor{myred}{rgb}{1,0.,0.3}

\usepackage[colorlinks=true,citecolor=myblue,linkcolor=myred]{hyperref}

\def\be{\begin{equation}}
\def\ee{\end{equation}}
\def\ba{\begin{align}}
\def\enda{\end{align}}
\def\bi{\begin{itemize}}
\def\ei{\end{itemize}}

\def\N{{\mathbb{N}}}

 \def\ee{\mathord{\rm e}}

\def\N{{\mathbb{N}}}

 \def\ee{\mathord{\rm e}}

\renewcommand{\ee}{{\rm e}}

\def\beq{\begin{equation}}
\def\beq{\begin{equation}}
\def\eeq{\end{equation}}
\def \ba{\begin{array}}
\def \ea{\end{array}}
\def \bea{\begin{eqnarray}}
\def \eea{\end{eqnarray}}
\def \bml{\begin{multline}}
\def \eml{\end{multline}}

 \newcommand{\ket}[1]{|#1\rangle}
 \newcommand{\bra}[1]{\langle #1|}
 
\newcommand{\bla}[1]{\left(#1\right)}
\newcommand{\blb}[1]{\left[#1\right]}

\newcommand{\bld}[1]{\left\langle#1\right\rangle}

\begin{document}

\title[Short Title]{Protecting a nuclear spin from a noisy electron spin in diamond}
\author{I. Cohen}
\affiliation{Racah Institute of Physics, The Hebrew University of Jerusalem, Jerusalem
91904, Givat Ram, Israel}
\author{T. Unden}
\affiliation{Center for Integrated Quantum Science and Technology, Universit\"{a}t Ulm, D-89069 Ulm, Germany}
\affiliation{Institut f\"{u}r Quantenoptik, Albert-Einstein Allee 11, Universit\"{a}t Ulm, D-89069 Ulm, Germany}
\author{F. Jelezko}
\affiliation{Center for Integrated Quantum Science and Technology, Universit\"{a}t Ulm, D-89069 Ulm, Germany}
\affiliation{Institut f\"{u}r Quantenoptik, Albert-Einstein Allee 11, Universit\"{a}t Ulm, D-89069 Ulm, Germany}
\author{A. Retzker}
\affiliation{Racah Institute of Physics, The Hebrew University of Jerusalem, Jerusalem
91904, Givat Ram, Israel}


\begin{abstract}
{Although a nuclear spin  is weakly coupled to its environment, due to its small gyromagnetic ratio, its coherence time is limited by the hyperfine coupling to a nearby noisy electron. Here, we propose to utilize continuous dynamical decoupling to refocus the coupling to the electron. If the random phase accumulated by the nuclear spin through the reduced coupling terms is sufficient small, we can increase the nuclear coherence time. Initially, we demonstrate this on a simple case with a two-level electron spin, while taking all relevant hyperfine coupling terms and noise terms  into account. We then extend the analysis to a nitrogen-vacancy center  in diamond having a three level structure.}
\end{abstract}
                                            
\maketitle

\section{introduction}
Nitrogen-vacancy (NV) centers in diamond are promising candidates for quantum information processing, due to their advanced quantum capabilities,  e.g., its perfect photostability and to its ground-state electron spin
properties, which combine a long coherence time \cite{Balasubramanian2009} and the ability to undergo
spin-sensitive optical transitions under ambient conditions \cite{Jelezko2004, Manson2006}. However, these electronic states are sensitive to noise and decoherence, giving rise to dephasing, which effectively limits the quantum processes in this system to $T_2$ duration. On the other hand, $^{13}C$ isotopes, in the vicinity of the NV center, possess a nuclear spin, which is more protected from the environment, and has a longer coherence time than the NV electron. Therefore, the nuclear spin can be utilized as a memory to store the quantum information. By employing the hyperfine coupling (HFC) between the proximate NV electron and $^{13}C$, one could conceivably exchange the electronic state for the nuclear one. However, due to a process combining the same HFC together with decaying electron states, the nuclear spin decoheres, and thus its performance as a quantum memory is limited. 

For nuclei in the vicinity of the NV, the HFC is stronger than the lifetime of the electron $ g_{HF} T_1 \gg1$, and thus, the nuclear spin dephasing time is equal to twice the electron lifetime $T_2^n=2 T_1$. To reduce this nuclear dephasing effect, the defect can be ionized in order to eliminate the noise term, like in \cite{Dreher,Pla,Saeedi2013} or with the aim to induce fast relaxation by photo-ionization to reduce the defect lifetime $T_1$ \cite{Lukin}.
In the former case, during each shortened $T_1$ time, the phase accumulated by the nuclear spin is a random variable $\delta \phi$ with a vanishing average $\bld{\delta \phi}=0$ and a variance $\bld{\delta \phi^2}=\bla{g_{HF} T_1}^2 \ll 1$, where $\bld{}$ is averaged over many experiments. In this case, the total random phase $\Delta \Phi$ obeys a random walk process, where each step takes $T_1$ time. After $N$ such steps, for which the variance of the total accumulated phase is $\bld{\Delta \Phi^2}/2 = N\bld{\delta\phi^2}/2 =1$, the phase is lost and we obtain the new nuclear coherence time $T_2^n=N T_1 = 2/(g_{HF}^2 T_1) \gg T_1$. Another options is to weakly irradiate of the NV and thus constantly initializing the NV in the $m_s = 0$ state \cite{Chen2017}.
However, in all these options limit the possibilities of utilizing the  defect electron for quantum information manipulation during this process. 

Instead of increasing the rate of decay, in this study we propose to reduce the HFC by utilizing continuous dynamical decoupling (CDD) \cite{Viola2003prl,Lidar2004pra,Fanchini2007pra,Kurizki}; namely, by driving the electron states continuously, to open protective energy gaps, which  compensate for the highest orders in the HFC.  As a result, the refocused HFC admits $g_{HF_{red}} T_1 \ll 1$, which decreases the noise on the nuclei considerably.

The ground state of the NMV center, e.g., NV center, forms a lambda system composed of electron spin sublevels with ms$=0, \pm1$. Therefore, to give a better intuition of the NV-$^{13}C$ problem above, we first consider a simpler case where the electronic degrees of freedom is a two level system. The two-level case is very relevant for current experiments involving spin one-half electron defects on the surface of the diamond \cite{Romach2015,Kim2015}, whose hyperfine coupling with the $^{13}C$ damages the nuclear coherence.


\section{Two-level electron system}
\begin{figure}[h]
\begin{center}
\includegraphics[width=0.45\textwidth]{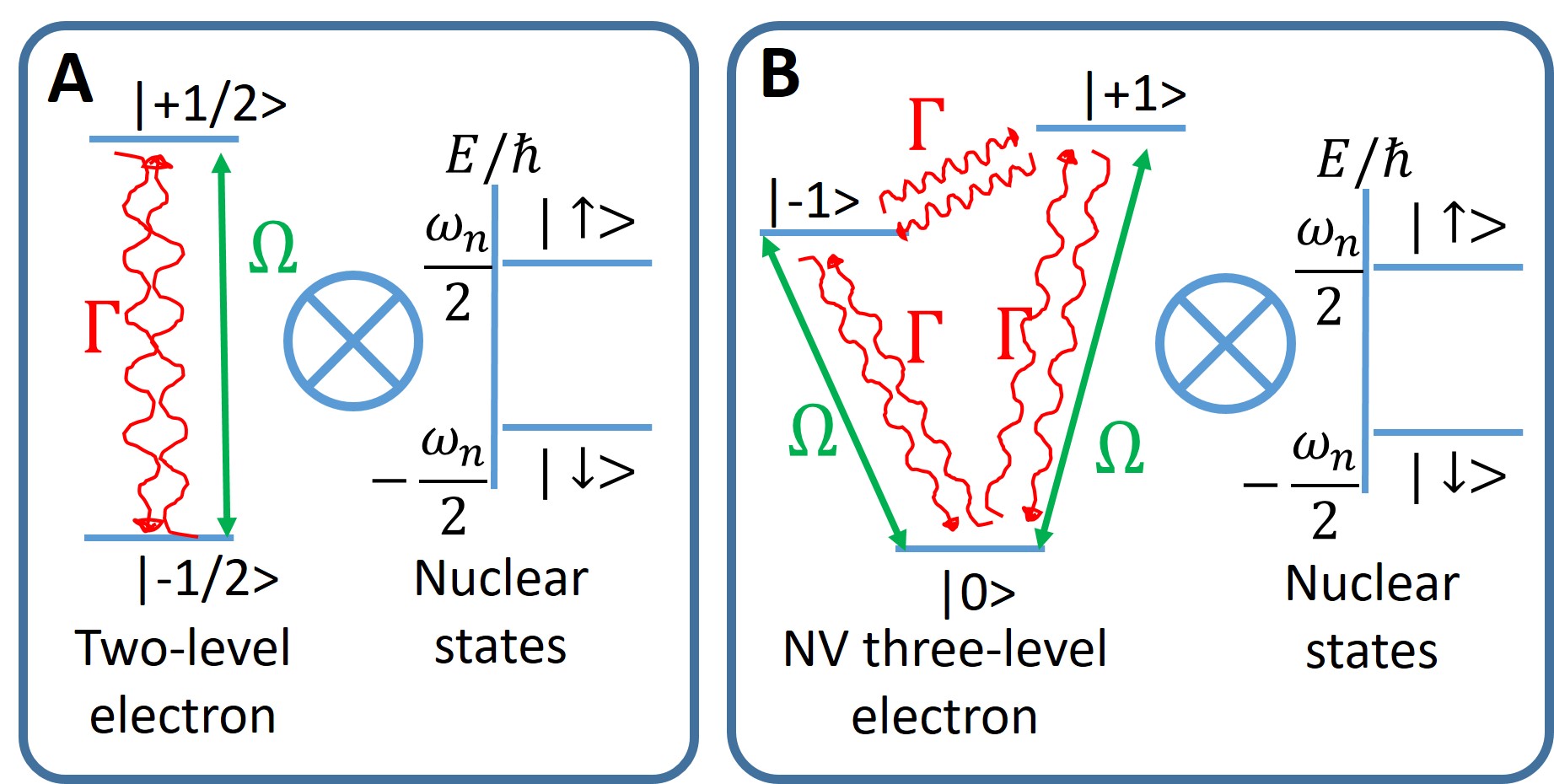}
\end{center}
\caption{{\bf Electron and nuclear spins.} (A). The electronic states comprise a two-level system subjected to random flip noise, $\Gamma$ (red arrows), caused by thermal phonons. (B) The NV electronic states establish a lambda system that is sensitive to decaying $\Gamma$ (red arrows). Due to HFC, the nuclear spin is coupled to the decaying electronic states, and therefore, the nuclear spin decoheres. By driving the electronic system resonantly with $\Omega$ (green arrow), we can reduce the HFC and the nuclear states are decoupled from decoherence. 
}
\label{energy_levels}
\end{figure}

For a simple case where the electron states constitute a two-level system, like a P1 center or a dangling bond on the surface, the electron flips randomly between the two levels with rate $\Gamma$ (fig. \ref{energy_levels}A). At room temperature, the lifetime of the electron spin is $T_1\sim1$ ms.  As a result of the HFC, the electron 
influences the nuclear states, and as a consequence, the nuclear spin dephases via the electronic decaying processes.

The system is described by the following Lindbladian 
\beq
\dot{\rho}= - i\blb{H,\rho}+L_e\bla{\rho},
\label{Lind}
\eeq
with the Hamiltonian in the rotating frame of the bare energy gap
\bea
H=\delta S_z + \omega_n I_{z}+g_\parallel S_{z} I_{z} + g_\perp S_{z} I_{x}, 
\label{H0}
\eea
and the electron decaying term
\beq
L_e\bla{\rho}=\frac{\Gamma}{2}\sum_{\alpha \neq \beta}^{+\frac{1}{2},-\frac{1}{2}} \bla{2\sigma_{\alpha,\beta}\rho \sigma_{\beta,\alpha}-\rho \sigma_{\beta,\alpha}\sigma_{\alpha,\beta}- \sigma_{\beta,\alpha}\sigma_{\alpha,\beta}\rho},
\eeq
with $|\delta|\approx (2\pi)100$ KHz, $\omega_n \approx(2\pi)100$ kHz, and $g_\parallel ,g_\perp \approx (2\pi)10$ kHz being a random detuning, the nuclear spin Larmor frequency, and the parallel and perpendicular HF coupling respectively; $\sigma_{i,j}=\ket{i}\bra{j}$ indicates the transitions between the electron states $i, j\in \{+1/2,-1/2\}$, $S_{\alpha}$ and $I_\alpha$ are Pauli matrices in the $\alpha^{th}$ direction normalized by a factor of 2, which describe the two-level electron and nuclear spins respectively. The random detuning term $\delta S_z$ describes a drift in the external magnetic field or a random magnetic field caused by unpolarized nuclear spins. In the Hamiltonian (Eq. \ref{H0}), we omit other HFC terms, e.g., the flip-flop $g_{ff}\bla{S_xI_x +S_yI_y}$, which are fast oscillating due to the very large energy difference between the electron and the nuclear spins. 


When the HFC is stronger than the electron decay rate $\Gamma \ll g_\parallel,g_\perp$, the nuclear spin dephasing time is limited by the electron life time $T_2^n=2 T_1$. To this end, we propose to reduce the HFC by applying CDD on the electron states. In CDD, a continuous driving field is applied resonantly, such that a protective energy gap is opened in the dressed state basis, to compensate for perpendicularly oriented noisy terms having sufficient long correlation times. We demonstrate here that CDD can also be used to refocus the HFC. By resonantly driving the $\ket{+1/2} \leftrightarrow \ket{-1/2}$ transition with Rabi frequency $\Omega\gg g_\parallel,g_\perp$ the Hamiltonian (Eq. \ref{H0}) reads
\beq
H=\delta S_z+{\omega_n} I_{z}+g_\parallel S_{z} I_{z}+g_\perp S_{z} I_{x}+{\Omega} S_x.
\label{protect}
\eeq
Now, during $T_1$ interval, before the electronic quantum jump occurs, the random phase accumulated by the nuclear spin fast oscillates, and we are left with reduced nuclear random phases $\sqrt{\bld{{\delta\phi_\parallel }^2}} = g_\parallel /\Omega \ll1$, and $\sqrt{\bld{{\delta\phi_\perp}^2}} = g_\perp/\Omega \ll1$. In addition, the total nuclear spin's phase $\Delta\Phi_\alpha (N)=\sum_{i=1}^N \delta \phi_\alpha(i)$, for $\alpha=\perp,\parallel$, are sums of independent random variables with vanishing averages. This yields a random walk, in which each $i^{th}$ step takes $T_1$ time. According to the central limit theorem, for a large $N$, this random process can be described by a normal distribution $\Delta\Phi_\alpha (N) \sim\N\bla{0,N \bld{{\delta \phi_\alpha}^2}}$. The new nuclear coherence time would be determined when 
\beq
\bld{e^{-\Delta\Phi_\alpha(N)}}=e^{-\frac{N \bld{{\delta \phi_\alpha}^2}}{2}}=\exp\bla{-1},
\label{coherence}
\eeq
for $\alpha=\perp,\parallel$, and where $\bld{}$ is averaged over a large number of experiments, and the left equality is a Gaussian identity. In this way, we obtain an increased nuclear coherence time $ {T_2^n}_\alpha =NT_1= 2 T_1 /\bld{{\delta\phi_\alpha}^2} \gg T_1 $. 

Although the first order of the HFC is refocused, the nuclear coherence time may not be increased, since there may still be a large contribution of higher orders of the HFC; namely an effective coupling  through which the nuclear spin might decohere. To simplify the derivation of the effective coupling term, we transform to the electron dressed state basis where $S_x \rightarrow F_z =\bla{ +\ket{+}\bra{+} -\ket{-}\bra{-}}/2$, $S_z \rightarrow -F_x$, $S_y \rightarrow F_y$, and the states $\ket{\pm 1} \rightarrow \bla{\ket{+}\pm \ket{-}}/\sqrt{2}$. In the rotating frame of both the bare energy structure of the nuclear spin ${\omega_n} I_z$ (fig. \ref{energy_levels}), and the electron dressed state energy ${\Omega} F_z$, we obtain 
\begin{multline}
H_I= 
 -\frac{\delta}{2} \bla{F_+e^{i \Omega t} + F_-e^{-i \Omega t} } \quad\quad\quad\quad\quad\quad\quad\quad\quad\\
-  \frac{g_\parallel}{2} \bla{F_+e^{i \Omega t} + F_-e^{-i \Omega t} } I_z \quad\quad\quad\quad\quad\quad\quad\\
- \frac{g_\perp}{4} \bla{F_+e^{i \Omega t} + F_-e^{-i \Omega t} } \bla{I_+e^{i \omega_n t} + I_-e^{-i \omega_n t} }.
\label{H_I}
\end{multline}
In the second order of perturbation theory, assuming $\Omega \gg \omega_n,\delta, g_\parallel,g_\perp$, we obtain the effective Hamiltonian which contains the A.C Stark shifts of both the electron and nuclear spins, in addition to an effective coupling term: 
\beq
H_{AC_{e}}=\bla{\frac{ g_\parallel^2 +\delta^2}{2\Omega} +\frac{g_\perp^2}{16} \bla{\frac{1}{\Omega-\omega_n}+\frac{1}{\Omega+\omega_n}  }  }F_z
\label{electron_AC}
\eeq

\beq
H_{AC_{n}}=\frac{g_\perp^2}{16}  \bla {\frac{1}{\Omega-\omega_n}-\frac{1}{\Omega+\omega_n}  }   I_z 
\label{nuclear_AC}
\eeq

\beq
H_{{coup}}= \frac{g_\parallel \delta}{\Omega}F_z I_z.
\label{eff_coupling}
\eeq
In this way, the effective coupling term can be reduced to $g_{eff}= g_\parallel \delta/{\Omega} \ll g_\parallel$, through which the nuclear spin looses its coherence. Assuming that the effective coupling is sufficient low such that during $T_1$ interval, the phase accumulated by the nuclear spin is small $\sqrt{\bld{{\delta \phi_{eff}}^2}} = g_{eff}T_1\ll 1$. As explained above (Eq. \ref{coherence}), we obtain an increased nuclear coherence time $ {T_2^n}_{eff} =NT_1= 2/{g_{eff}}^2 T_1 \gg T_1$. The resulted nuclear coherence time is thus determined by the strongest decoherence process: $T_2^n=\bla{1/{T_2^n}_\parallel+1/{T_2^n}_\perp+1/{T_2^n}_{eff}}^{-1}$.


Although the effective coupling term is reduced, it may still be larger than the electron lifetime $g_{eff}T_1>1$, negating the protection of the nuclear spin. We therefore propose to reduce 
the $F_z I_z$ coupling term by utilizing concatenated CDD \cite{Cai2012njp}; namely, by opening another electron energy gap, in a perpendicular direction to the dressed-state basis. There are three ways to open this protective energy gap: (1.) modulating the magnetic field in the $z$ direction, with a modulation frequency resonant with the dressed state energy gap, namely, the Rabi frequency of the first driving field \cite{Itsik2015njp}; (2.) utilizing the time-dependent detuning technique by adding a phase modulation $\bla{\Omega_2/\Omega}\sin \Omega t$ to the first driving field \cite{Itsik2016fp}; (3.) driving the electron bare states $\ket{+1/2}\leftrightarrow \ket{-1/2}$ with two opposite  detunings $\pm\Omega$, having the same Rabi frequency $\Omega_2$ \cite{Cai2012njp}. 
\begin{figure}[h]
\begin{center}
\includegraphics[width=0.45\textwidth]{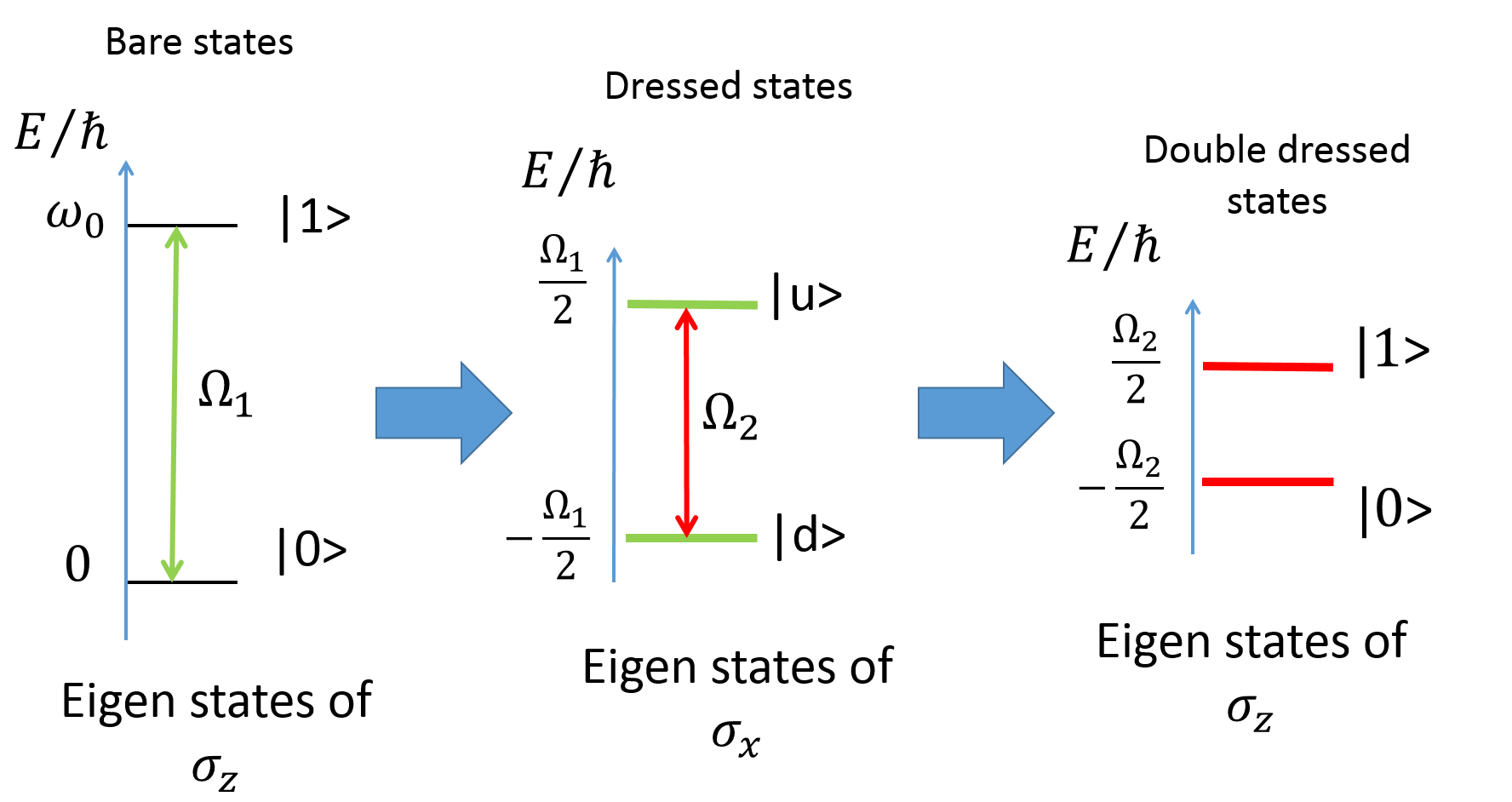}
\end{center}
\caption{{\bf Concatenated CDD.} 
(Left) By applying a driving field (green) on resonance with the bare energy gap we move to the dressed states (Center), where we open a protective energy gap to refocus the HFC. Then, using one of the three ways (red) we open a second protective energy gap (Right)  in the double dressed state basis, to further reduce the HFC, resulting in a better protection of the nuclear spin.
}
\label{second}
\end{figure}

Either of the first two options contributes another term to the Hamiltonian (Eq. \ref{protect}) in the interaction picture
\beq
H_{d2}=2\Omega_2 S_z \cos \bla{\Omega t}=-2\Omega_2 F_x \cos \bla{\Omega t }.
\eeq
Assuming $g_{eff}\ll\Omega_2 \ll \Omega$, then this additional term becomes a second protective energy gap in the rotating frame of the dressed state energy gap, together with fast oscillating terms 
\beq
H_{d2_I}=-{\Omega_2}{F_x} -\frac{\Omega_2}{4}\bla{F_+ e^{2i\Omega t} +h.c },
\label{second_gap1}
\eeq
The third option produces 
\beq
H_{d2}=2\Omega_2 S_y \cos  \bla{\Omega t}=2\Omega_2 F_y \cos  \bla{\Omega t },
\eeq
resulting in a similar protective energy gap together with the fast rotating terms
\beq
H_{d2_I}={\Omega_2} {F_y }+\frac{\Omega_2}{4}\bla{+i F_+ e^{2i\Omega t} -h.c }.
\label{second_gap2}
\eeq
These fast oscillations (second terms of Eq. \ref{second_gap1}, \ref{second_gap2}),  can usually be neglected since they average to zero. However, although at this point, these terms are ignored, we show later they can result in effective coupling in a higher order of perturbation. By all three options, the reduced nuclear random phase during the $T_1$ interval is $\sqrt{\bld{{\delta\phi_{\Omega_2}}^2}} = g_{eff}/\Omega_2 \ll1$, giving rise to an improved nuclear coherence time ${T_2^n}_{\Omega_2}= 2T_1/\bld{{\delta\phi_{\Omega_2}}^2}$. Similarly to the above explanation, if there is a detuning from the dressed state resonance $\delta_2 F_z$ we are still left with a further reduced coupling term
\beq
H_{eff_2}=g_{eff_2}F_\theta I_z = \frac{g_{eff} \delta_{2}}{\Omega_2}F_{y,x} I_z.
\label{coupling_eff_2}
\eeq 
We assume that the random phase accumulated by the nuclear spin during the $T_1$ interval, as a result of this term, is small $\sqrt{\bld{{\delta\phi_{eff_2}}^2}}=g_{eff_2}T_1 \ll 1$, thus the resulting nuclear coherence time is ${T_2^n}_{eff_2}=2/g_{eff_2}^2 T_1 \gg T_1$. 

Note that the detuning term $\delta_2 F_z$ might originate from drifts in the first Rabi frequency, or from the A.C Stark shift in Eq.\ref{electron_AC}. Assuming the drifts are small, we can still consider the contribution of the bare detuning $\delta$ originating from the unpolarized nuclear spins, as is behaves as $\delta^2$, and thus is invariant to the unpolarized nuclear spin state. 

Even if we eliminate the detuning from the dressed state energy $\delta_2=0$, there is still a higher order coupling. Using the effective Hamiltonian description, by taking the time independent terms of the third order of Magnus expansion, 
\begin{multline}
H_{eff_3}=\frac{1}{6 t}\int_0^t dt_1\int_0^{t_1} dt_2\int_0^{t_2} dt_3 \cdot\\
\bla{\blb{H(t_1), \blb{H(t_2),H(t_3)} }+ \blb{H(t_3), \blb{H(t_2),H(t_1)} } }
\end{multline}
we see that there is an additional effective coupling term, originating from a combination of the fast oscillating terms of the second energy gap (second terms of Eq. \ref{second_gap1} \ref{second_gap2}) that rotate with $2\Omega$  and the first two lines of Eq. \ref{H_I} that rotate with $\Omega$.
\beq
H_{eff_3}=g_{eff_3}F_x I_z =\frac{g_\parallel  \Omega_2\delta}{2\Omega^2} F_x I_z.
\eeq
Now, $g_{eff_3} \ll g_{eff}$, and if the random phase accumulated by the nuclear spin during $T_1$ time $\sqrt{\bld{{\delta\phi_{eff_3}}^2}}=g_{eff_3} T_1 \ll1$, the increased nuclear coherence time is ${T^n_2}_{eff_3}=2T_1/\bld{{\delta\phi_{eff_3}}^2}$. The resulted nuclear coherence time is thus determined by the strongest decoherence process: 
\beq
T_2^n=\bla{\frac{1}{{T_2^n}_\parallel}+\frac{1}{{T_2^n}_\perp}+\frac{1}{{T_2^n}_{\Omega_2}}+\frac{1}{{T_2^n}_{eff_2}}+\frac{1}{{T^n_2}_{eff_3} }}^{-1}.
\label{total_T2_2}
\eeq
In our simulation, we see that according to our assumed parameters, the nuclear coherence time is determined by ${T^n_2}_{eff_3}$, which is the strongest decoherence process (Fig. \ref{sim}).

\section{Introducing noise}
As the nuclear spin is decoupled from its environment, the sources of noise are associated with the electron spin. The main ones are the ambient magnetic field fluctuations, induced by  substitutional nitrogen impurities in diamond (P1 centers) \cite{P1_defect}, and the Rabi frequency fluctuations of the driving fields. The magnetic noise $\delta B\bla{t}S_z$ and the Rabi frequency noise $\delta \Omega\bla{t}S_x$ can be described by the Ornstein-Uhlenbeck (OU) process \cite{OU1,OU2} with a zero expectation value, $\bld{\delta O\bla{t}}=0$,
and a correlation function $\bld{\delta O\bla{t}\delta O\bla{t'}} =\frac{C_O\tau_O}{2}e^{-\left|t-t'\right|/\tau_O}$, where $C_O$ is the diffusion constant and $\tau_O$ is the correlation time of the magnetic noise or the Rabi frequency noise having $O=B$, and  $O=\Omega$ respectively.  The OU process is simulated by an exact algorithm \cite{OU3}, according to
\begin{equation}
\delta O(t+\Delta t)=\delta O(t)e^{-\frac{\Delta t}{\tau}}+n\sqrt{\frac{C_O\tau_O}{2}\bla{1-e^{-\frac{2\Delta t}{\tau_O}}}},
\end{equation}
where $n$ is a unit Gaussian random number. 

Regarding the magnetic noise we assume a correlation time of  $\tau_B=25\:\mu s$ \cite{P1_defect}, and an amplitude of $\delta B(t) \sim \sqrt{\frac{C_B \tau_B}{2}} \ll \Omega$. Since the Rabi frequency is greater than the correlation time of the magnetic field noise $\Omega\tau_B\gg1$, we can still obtain the effective Hamiltonian that gives rise to an effective coupling term (Eq. \ref{eff_coupling}):
\beq
H_{{coup}}= \frac{g_\parallel \blb{\delta+\delta B\bla{t}}}{\Omega}F_z I_z.
\eeq
Therefore, during the correlation time $\tau_B$, meaning the time in which the amplitude of the magnetic noise varies, we gain a random phase of $\sqrt{\bld{{\delta\phi_{\delta B(t)}}^2}}=g_\parallel \delta B\bla{t}\tau_B/\Omega $. If this random phase is large $\sqrt{\bld{{\delta\phi_{\delta B(t)}}^2}}>1$, the protection of the nuclear spin is damaged, which means that we have to employ the second driving field, and have to assume that $\Omega_2\tau_B\gg1$. In this case the random phase accumulated due to the magnetic noise is $\sqrt{\bld{{\delta\phi_{{\delta B(t)}_{2}}}^2}}=g_\parallel \delta B\bla{t}/\Omega\Omega_2 \ll1 $, resulting in a nuclear coherence time of ${T_2^n}_{\delta B(t)}=2\tau_B/\bld{{\delta\phi_{{\delta B(t)}_{2}}}^2}$.

With respect to the Rabi frequency noise, if we assume a correlation time of $\tau_{\Omega}=100\:\mu s$, and an amplitude error of $\delta \Omega\bla{t}=0.005\Omega$, then the diffusion constant is given by $C_{\Omega}=2{\delta \Omega\bla{t}}^2 / \tau_{\Omega}$. The most dominant effect of this noise term is the detuning we obtain when opening the second energy gap; namely, $\delta_2=\delta \Omega\bla{t}$ in Eq. \ref{coupling_eff_2}. During the correlation time of the Rabi frequency noise, we obtain a random phase of $\sqrt{\bld{{\delta\phi_{\delta\Omega(t)}}^2}}=g_{eff} \delta\Omega(t) \tau_\Omega/\Omega_2 $, due to this coupling term, giving rise to a nuclear coherence time of ${T_2^n}_{\delta \Omega(t)}=2\tau_\Omega/\bld{{\delta\phi_{\delta\Omega(t)}}^2}  $. 

\begin{figure}[h]
\begin{center}
\includegraphics[width=0.45\textwidth]{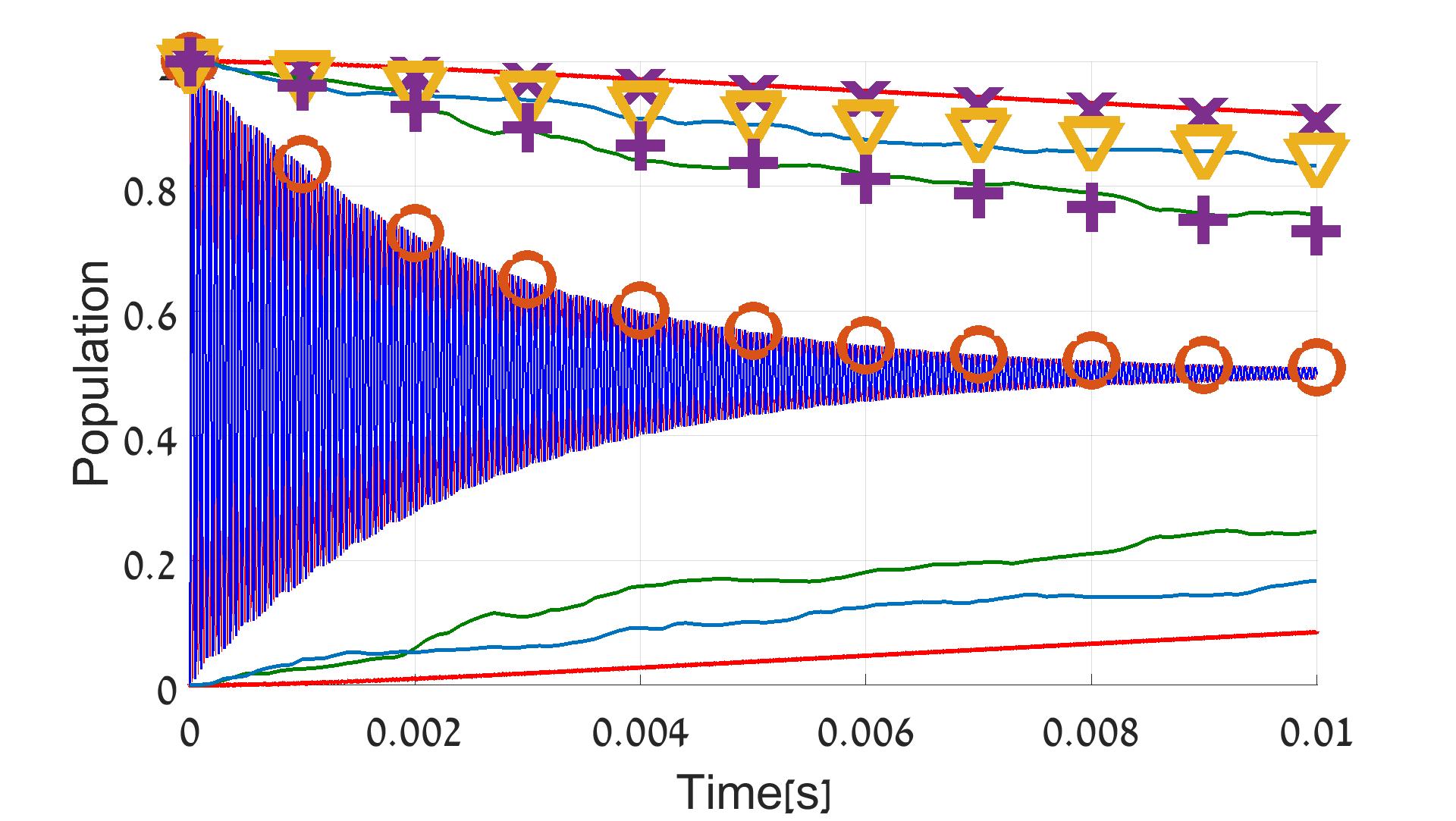}
\end{center}
\caption{{\bf Nuclear spin dephasing.} We simulate the system of a nuclear spin coupled to a decaying electron having either two or three levels, with the following parameters: $\omega_n=(2\pi)100$ kHz, $g_\parallel=(2\pi)40$ kHz, $g_\perp=(2\pi)20$ kHz, $\delta=(2\pi)100$ kHz, $T_1=1.25$ ms. We then measure the population of the nuclear spin: (Blue oscillations) Without any protection we observe the oscillations in frequency $g_\parallel$ in both systems. (Red circles) The nuclear decoherece is fitted to $|\bra{+}_I U_{sim} \ket{+}_I|^2 = 0.5\blb{1+\exp\bla{-t/T_2^n}}$ with a nuclear coherence time of $T_2^n=2 T_1$. (Dotted red curves) With a protection of $\Omega=(2\pi)4$ MHz, $\Omega_2=\Omega/17$, and by adjusting $\Delta\Omega$ in Eq. \ref{coupling2}, both systems behave exactly the same, where the lower(upper) dotted curve is the population of $\ket{+}_I$ ($\ket{-}_I$) states. (Purple exes) We fit to a decay rate of $T_2^n={T_2^n}_{eff_3}=0.047$ s, which agrees with Eq. \ref{total_T2_2}. (Solid curves) We introduce the magnetic and Rabi frequencies noises, with the following parameters: $\delta B(t)=(2\pi) 50$ kHz, $\tau_B=25\,\mu$s (thus ${T_{2}^e}*=4\,\mu$s), $\delta \Omega(t) /\Omega=\delta \Omega_2(t) /\Omega_2=\delta \Delta\Omega(t) /\Delta\Omega=0.005$, $\tau_\Omega=\tau_{\Omega_2}=\tau_{\Delta\Omega}=100\,\mu$s. Here, we notice a deviation between the two systems. In both systems the main decoherence contributions originate from both ${T_2^n}_{eff_3}$, and the Rabi frequency noise ${T_2^n}_{\delta\Omega(t)}$. The Rabi frequency noise of the NV three-level system (Green solid curves) s is twice as much as the Rabi frequency noise of the two-level systems (Blue solid curves). Therefore, the coherence time of the two levels system is ${T_2^n}_{\delta\Omega(t)}=0.07$ s, which is four times longer than the coherence time of the NV system ${T_2^n}_{\delta\Omega(t)}=0.0175$. The combination of the Rabi noise together with the ${T_2^n}_{eff_3}$ giving rise to $T_2^n=0.028$ s and $T_2^n=0.0128$ s, as fitted (Yellow triangles) and (Purple pluses), for the two-level system and the NV system respectively.
}
\label{sim}
\end{figure}

The Rabi frequency noise of the second driving field does not couple to the HFC, thus it does not interfere with the nuclear protection. In our simulation, which includes noisy terms, the Rabi frequency noise has the strongest decoherence effect on the nuclear spin, together with the decoherence process of ${T^n_2}_{eff_3}$ (Fig. \ref{sim}).

\section{NV three-level system}
Progressing from the simple case of two-level electron spin, the NV electron states constitute a triplet lambda-system, which randomly jumps from one state to another with rate $\Gamma$ (fig. \ref{energy_levels}). The NV-$^{13}C$ system is described by a similar Lindbladian (Eq. \ref{Lind}) with a similar Hamiltonian (Eq.\ref{H0}), in which the only differences are that $S_\alpha$ describes the electron triplet (spin-one) angular momentum matrix in the $\alpha^{th}$ direction, and the electron decaying term involves all decay processes
\beq
L_e\bla{\rho}=\frac{\Gamma}{2}\sum_{\alpha\neq\beta}^{-1,0,+1} \bla{2\sigma_{\alpha,\beta}\rho \sigma_{\beta,\alpha}-\rho \sigma_{\beta,\alpha}\sigma_{\alpha,\beta}- \sigma_{\beta,\alpha}\sigma_{\alpha,\beta}\rho}.
\eeq
Although the decaying processes are not isotropic, we assume isotropy here, to obtain a simpler theoretical analysis, which agrees with the simpler two-level electron case. Of course, our protection scheme can be implemented on the non-isotropic decaying case.

Employing the CDD technique to refocus the HFC in the NV three-level system is done in the following way. We suggest to drive both $\ket{\pm 1} \leftrightarrow \ket{0}$ transitions simultaneously with the same Rabi frequency $\sqrt{2}\Omega$; thus we drive the spin-one $\Omega S_x$ transition continuously. In addition, we introduce a small difference $\Delta \Omega$ in the Rabi frequencies between the transitions $ \ket{+1} \leftrightarrow \ket{0}$ and $ \ket{-1} \leftrightarrow \ket{0}$, therefore, the Hamiltonian reads
\beq
H=\delta S_z+{\omega_n} I_{z}+g_\parallel S_{z} I_{z}+g_\perp S_{z} I_{x}+\Omega S_x + \Delta \Omega \bla{\sigma_{+1,0} + h.c}.
\label{protect}
\eeq
Similar to the derivation above, the HFC is refocused thanks to the dressed state energy gap $\Omega S_x$. During the $T_1$ interval, the random phase accumulated by the nuclear spin fast oscillates, and we are left with reduced nuclear random phases $\sqrt{\bld{{\delta\phi_\parallel}^2}} = g_\parallel /\Omega \ll1$, and $\sqrt{\bld{{\delta\phi_\perp}^2}} = g_\perp/\Omega \ll1$. Therefore, we obtain an increased nuclear coherence time $ {T_2^n}_\alpha =NT_1= 2 T_1 /\bld{{\delta\phi_\alpha}^2} \gg T_1 $, for $\alpha=\parallel,\perp$.

Through the effective coupling terms, the nuclear spin might decohere. To calculate these effective coupling terms we first move to the electron dressed state basis where $S_x \rightarrow F_z = +1\ket{u}\bra{u} + 0 \ket{D}\bra{D} -1\ket{d}\bra{d}$, $S_z \rightarrow -F_x$,  and $S_y \rightarrow F_y$, and the states $\ket{+1} \rightarrow \bla{\ket{u}+\ket{d}}/2 -\ket{D}/\sqrt{2}$, and $\ket{0} \rightarrow \bla{\ket{u}-\ket{d}}/\sqrt 2$. Then, we move to the interaction picture with respect to the bare energy structure of the nuclear spin $\omega_n/2 I_z$ (fig. \ref{energy_levels}), and the electron dressed state energy $\Omega F_z$. Thus we obtain the same terms of Eq. \ref{H_I} of the previous case, together with the following
\beq
H_{I_{additional}}=\Delta \Omega \bla{\frac{1}{\sqrt{2}} F_z  -\frac{1}{2\sqrt{2}}\bla{\tilde{F}_+ e^{i\Omega t}+\tilde{F}_- e^{-i\Omega t}} }.
\eeq
where $\tilde{F}_+ = \sqrt{2}\bla{\ket{u}\bra{D} - \ket{D}\bra{d}}$, and ${F}_+ = \sqrt{2}\bla{\ket{u}\bra{D} + \ket{D}\bra{d}}$.

In the second order of perturbation theory we obtain an effective Hamiltonian similar to that in the two-level system (Eq. \ref{electron_AC},\ref{nuclear_AC},\ref{eff_coupling}), with the only additional coupling term being:


\beq
H_{{coup_2}}=-\bla{\frac{g_\perp^2}{8}   \bla{\frac{1}{\Omega-\omega_n}-\frac{1}{\Omega+\omega_n}  }  -\frac{3g_\parallel \Delta\Omega}{2\sqrt{2}\Omega}}  \bla{F_z}^2   I_z 
\label{coupling2}
\eeq

Although these coupling terms are reduced, they might be larger than the electron lifetime $g_{eff}T_1>1$, resulting in a non effective protection of the nuclear spin. As shown above, we can utilize the concatenated CDD to further reduce these effective coupling terms. Following the above derivation of utilizing the the concatenated CDD to refocus Eq. \ref{eff_coupling}, for a three level electron system, we obtain the same results. Note that refocus Eq. \ref{coupling2} might be problematic, since an $\bla{F_z}^2$ term cannot be refocused using CDD; however, by adjusting the Rabi frequency difference $\Delta\Omega$ this coupling term can be completely removed.


\section{Introducing noise}
The same noise treatment as that in the two-level scenario is also valid here, as long as the Rabi frequency noise is correlated. Namely, the Rabi frequencies driving both $\ket{\pm 1} \leftrightarrow \ket{0}$ transitions should originate from the same source, and so be correlated. Note that if these Rabi frequencies are not correlated, this will introduce a very large noisy Rabi frequency mismatch $\Delta \Omega(t) \sim 0.005\Omega$, resulting in a heavily noisy Eq. \ref{coupling2} that can not be refocused continuously. We therefore assume that the noise in $\Delta \Omega(t) \sim 0.005 \Delta \Omega$ originates only from the additional drive with $\Delta \Omega \ll \Omega$, and the random phase accumulated by the nuclear spin $\delta\phi_{\delta\Delta\Omega(t)}= 3 g_\parallel \delta\Delta\Omega(t)/\sqrt{2}\Omega \ll 1$ is negligible. 

Importantly, because of the three-levels, the noisy terms give rise to double random phases compared to the two-level system, as our simulation shows (Fig. \ref{sim}).

\section{Summary}
In this manuscript we present a novel method to protect the nuclear spin from the noise inflicted by a nearby electron. Instead of reducing the lifetime of the electron, we employ a CDD technique  to reduce the HFC. To this end, we open two protective energy gaps by driving the electron states, whether these comprise two or three energy levels. We have derived the refocused coupling terms to the highest contributing order, through which the nuclear spin possesses a reduced dephasing, and thus, we can calculate the increased nuclear coherence time, which scales as $T_2^n \propto \bla{\Omega/g_\parallel \delta}^2$. For our assumed parameters, we have managed to increase the nuclear coherence time by an order of magnitude. However, for lower HFCs and larger protecting energy gaps, we can obtain a further decoupled nuclear spin. It is noteworthy that an extension of the method we present could be used to decouple the nuclei from the NV center while polarizing the NV center or while performing a quantum non demolition measurement of different nuclei \cite{QND1,QND2}. However, in this case both the ground state manifold and the excited state should be driven.

It is also important to note that one could implement the pulse dynamical decoupling technique rather than the continuous approach that was shown in this paper. 

\section{Acknowledgements}
 A. R. acknowledges the support of the Israel Science Foundation(grant no. 1500/13), the support of the European commission, EU Project DIADEMS.  
This project has received funding from the European Union a��s Horizon 2020 research and innovation programme under grant agreement No 667192 Hyperdiamond and Research Cooperation Program and DIP program (FO 703/2-1).


\end{document}